\documentstyle[twoside,fleqn,espcrc2]{article}
\newcommand{\AmS}{{\protect\the\textfont2
  A\kern-.1667em\lower.5ex\hbox{M}\kern-.125emS}}

\title{Branes Probing Black Holes.
}

\author{Juan M. Maldacena\address{Lyman Laboratory of Physics, Harvard
University, \\
Cambridge, MA 02138, USA}%
        \thanks{Expanded version of the talk presented at Strings97, 
June 1997.}
       }

\def\be{\begin{equation}}
\def\ee{\end{equation}}
\newcommand{\la}[1]{\label{#1}}
\def\ba{\begin{array}}
\def\ea{\end{array}}
\newcommand{\refnew}[1]{(\ref{#1})}

\def\sq2{\sqrt{2}}

\def\s42{ 2^{-{1\over 4} } }

\def\[{\left [}
\def\]{\right ]}
\def\({\left (}
\def\){\right )}

\input epsf.tex

\begin{document}

\bibliographystyle{unsrt}

\begin{abstract}

We consider a brane moving close to a large 
number of coincident  branes. We compare the
calculation of the effective action using the gauge theory 
living on the brane 
and the calculation using the supergravity approximation.
We discuss some general features about the correspondence between
large $N$ gauge theories and black holes.
Then we do a one loop calculation which applies for extremal 
and near extremal black holes. 
We comment on the expected results for higher loop calculations.
We make some comments on  the Matrix theory interpretation
of these results.

\end{abstract}


\maketitle

\section{Introduction}

D-branes are localized probes of the  spacetime  geometry
\cite{mdgm,douglasfivebrane,douglasseiberg,dkpp,dkps}.
When a D-brane probe is in the presence of other D-branes there are
massive open strings stretching between the probe and the other
branes. 
When these massive open strings are integrated out one obtains
an effective action for the massless fields representing 
the motion of the probe.
This can be  interpreted  as the action of a D-brane
moving on a nontrivial supergravity background.
In many cases one can find the exact supergravity
backgrounds in this fashion \cite{douglasfivebrane,douglasseiberg,dkps}.
Most of the backgrounds analyzed previously correspond to BPS supergravity
solutions. 
The calculation in  \cite{dps} of D-brane probes moving in a 
nonsingular extremal black hole background shows
agreement for the one loop term but the status of the two loop term
is not clear. 

We will first consider the field theory problem, including the
problem
of decoupling  the brane  theory from the whole string theory. 
We discuss some general properties of the effective action for
these gauge theories derived by using power counting arguments. 
We also do a one loop calculation in this gauge theory.

Then we consider the effective action of a probe D-brane moving 
close to a black D-p-brane. We discuss some cases where we can 
take the same limit as we took in the gauge theory case. 
We find  a limit where the black hole is well defined and we
can apply supergravity. 
We mention that this predicts certain results for the gauge theory.

We discuss the relation of these results with Matrix theory, in
particular, the relationship of black holes and Matrix theory.
We discuss the case of the D-twobrane in some detail, including the
appearance of a nontrivial strong coupling IR fixed point and its
black hole interpretation.

We finally consider near extremal black holes  and we compare the
one loop result obtained in both ways. 
 Since the background  is 
no longer BPS and supersymmetry is broken there is no reason for
the forces to cancel. Indeed there is a net force on a static probe.
We also calculate the $v^2$ force and we find agreement  with
supergravity
 in {\it all }
cases.
We compute the static force 
for a  D-brane  configuration with $Q_5$ D-fivebranes carrying also $Q_1$
D-onebrane
charge and some extra energy. We find precise agreement. The static force
agrees qualitatively in all the other  cases.
All one loop calculations of this type reduce to evaluating
 $F^4$ terms in the gauge theory.

\section{ The field  theory problem}

In this section we will consider a D-brane probe in the presence
of  some other D-branes. By integrating out the stretched 
open strings we  obtain an  effective 
action for the probe.
More concretely, we consider 
$N+1$ D-branes, $N$ of which sit at the origin ($r=0$)  and
the last is  the probe which sits at a distance $r$.
At low energies and for small separations the system is
described by a $p+1$ dimensional 
$U(N+1)$ Yang-Mills theory with 16 supersymmetries
broken down
to $U(N)\times
U(1)$ by the expectation value of an adjoint scalar which 
measures the distance from  the probe to the rest of the branes.
The fields  with one index in $U(N)$ and the other on $U(1)$ 
are massive, with a mass $ m = r/(2 \pi \alpha')$. 
If we integrate them out we get an effective action 
for the light degrees of freedom.
All  D-branes  have $p+1$ worldvolume dimensions, with $p \le 6$.

\vskip 1cm
\vbox{
{\centerline{\epsfxsize=2in \epsfbox{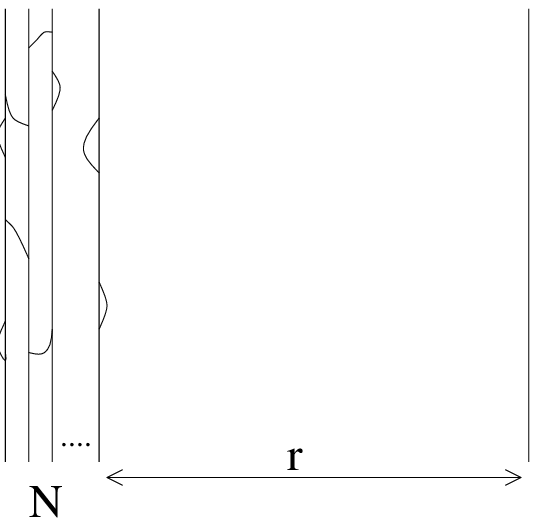}}}
{\centerline{\tenrm FIGURE 1: D-brane configuration.}}
{\centerline{ N D-branes
are sitting together and }}
{\centerline{ carrying low energy  excitations.}}
{\centerline{ The probe is separated 
by a distance $r$.}}
}
\vskip .5cm

The action of the Yang Mills theory is 
\be \label{action} \ba{rl} 
S_0 =& - { 1 \over g (2\pi)^p \alpha'^{p-3\over 2} } \int d^{1+p}x  \ \ 
{1\over 4} Tr[ F_{\mu\nu} F^{\mu\nu} ] \\
&+
{\rm fermions }~, \ea 
\ee
where $\mu,\nu$ are ten dimensional indices \cite{notesondbranes}.
Let us first discuss more precisely the sense in which \refnew{action}
describes
the dynamics of nearby branes more precisely. 
We can take the low energy limit of the string theory by taking
the energies to be fixed and letting $\alpha' \to 0$. In order for
the energies of the stretched strings to remain finite we should let
the separation of the branes $ r \to 0$ in such a way that
$A^I\equiv X^I \equiv r^I/\alpha' $ is fixed. The field $X^I$ is the field 
describing transverse displacements of the branes and has dimensions
of energy. In this way the integrand in \refnew{action} has no explicit
factors
of $\alpha'$. We should also, at the same time, make sure that the 
gauge theory coupling constant remains finite. This implies that
$g \alpha'^{p-3\over 2}$ is fixed. For $p<3$ this implies that 
$g \to 0$, for $p=3$ it does not impose any condition on $g$ and
for $p>3$ it implies that $g \to \infty$. So for $p>3$ we should
analyze the problem in a dual theory. For $p=4$ we have in this 
limit the (0,2) theory of coincident 5-branes \cite{andy} 
in M-theory compactified
on a circle of radius $R= g \sqrt{\alpha'}$. For $p=5$ we have to do a IIB 
S-duality transformation and we end up with NS fivebranes in the
limit that $\tilde g = 1/g \to 0$ and the coupling constant is
$\tilde \alpha' = g \alpha' ={\rm fixed}$. 
In this limit we obtain theories decoupled from the bulk 
\cite{seibergfive}.
For $p=6$ it is not clear whether the theory really decouples
from the bulk.
Notice that, in this limit, the probe brane is getting very close to the
other branes, at  a  distance  much smaller than
the string length. 
In summary, the limit which produces a decoupled brane theory is
\be \la{decouplbr}\ba{rl}
 ~&\alpha' \to 0~,~~~~~~\omega ={\rm fixed}~, \\ ~& g^2_{YM} = 
(2 \pi)^{p-2} g
 \alpha'^{p-3 \over 2 } = {\rm fixed } \\
  ~& X^I = {r^I \over \alpha'} = {\rm fixed }
\ea
\ee
With some abuse of the language we will call these ``gauge'' theories 
on the branes, with the understanding that in the $p=4,5$ cases we
are referring to the theories of coincident M- and NS-branes respectively.

Now let us consider the case where we have a nontrivial field strength
on the $N$ p-branes and we calculate the effective action for the
probe. This calculation will involve diagrams with insertions 
of the  field $A_\mu$ on the brane. 
We expect the result to be gauge invariant, so we expect that
$A_\mu$ only appears through the field strength.

This calculation will involve Feynman diagrams as shown in 
figure 2 which have some number $I$ of insertions of the gauge field
and some number of loops $L$. Since we have an $L$ loop diagram we
will
have a factor of $g_{YM}^{2L-2}$. The integrals in the diagram will
involve some massive particles which have a mass of the order
of $X \equiv r/\alpha'$ and this is the only scale that appears in
the integrals. Assuming that the integrals are convergent  we 
get the following behavior for the diagram with $L$ loops and $I$
insertions
\be\la{powerc}\ba{rl}
&g_{YM}^{2L-2} N^{L}{F}^{I}  { 1 \over X^{L(3-p) +2 (I  -2)} }  = \\
&~~~~~~~=  { F^2 \over g_{YM}^2 }
\left( g_{YM}^2 N F^2 \over X^{7-p} \right)^L  \left( F \over X^2 
\right)^{I - 2 L -2 }  \ea
\ee
We have concentrated  only in the leading $N$ behavior, there
could be contributions that are subleading in $N$. 
The term $F^I$ indicates some particular contraction of the gauge field
strength, both in the YM indices and the Lorentz indices.

\vskip 1cm
\vbox{
{\centerline{\epsfxsize=2in \epsfbox{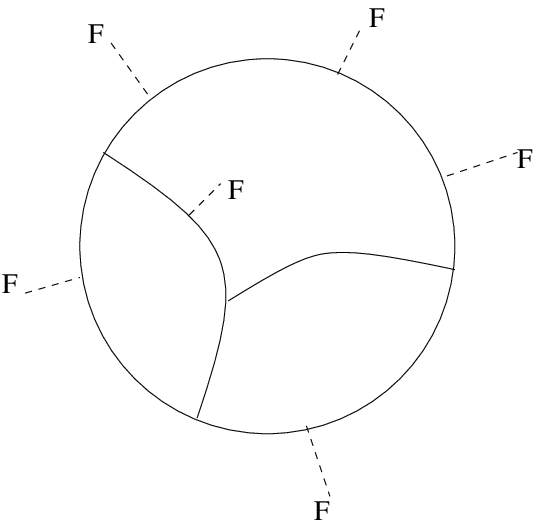}}}
{\centerline{\tenrm FIGURE 2: Typical Feynman diagrams.}}
}
\vskip .5cm

As an example we can consider the one loop contribution, this 
 can be extracted from the calculations in 
\cite{callancorrections,callanconstant} 
and it is a very simple generalization of the calculations
done by \cite{bachas,cakl,lima,dkps,makeenko}. We 
refer to those  papers for the details. 
We will calculate the leading order terms in the field strength 
$F$.
 We will assume that $F$ is slowly varying so that
we can neglect derivatives,  $DF$, as well as commutators, $[F,F]$. 
Commutators are small if covariant derivatives of the fields
are small since $ [ D , D ] F \sim [ F, F ] $.
A similar  approach was taken in \cite{tseytlindbi} to propose a form
for the non-abelian generalization of the Dirac-Born-Infeld action.
The diagrams with less than four external lines are zero so that
the first nonvanishing contribution is \cite{yoprobes}
\be\la{result}\ba{rl}
S_1 = & { c_{7-p}\over 8 ( 2\pi )^p  X^{7-p} } 
\int d^{1+p } x \ \left\{ Tr [ F_{\mu \nu} F^{\nu \rho} F_{\rho \sigma }
F^{\sigma \mu }]  - \right.\\
& \left. {1 \over 4 } Tr [ F_{\mu \nu}F^{\mu \nu } 
F_{\rho \sigma }F^{\rho \sigma } ] \right\}~, \ea
\ee
where $\mu , \nu$ are again ten dimensional indices and
$c_q$ is the numerical constant 
\be\la{defcp}
c_q = (4  \pi)^{{q  \over 2 } -1 } \Gamma( { q \over 2 })~ . 
\ee
We see that there are two terms 
with a structure that  is independent of the
dimensionality of the brane. This structure is the same
as the structure of the $F^4$ terms in the Dirac-Born-Infeld 
action \cite{tseytlindbi,tseytlinmet}.
The case of branes moving with constant velocity corresponds
to $F_{0i} =v^i$ and  we get 
\cite{bachas,cakl,dkps}  
\be\la{vfourth}
S = { c_{7-p} N \over 8 (2 \pi)^p  r^{7-p} } \int d^{p+1}x ( v^2 )^2~.
\ee

We also see that if $F$ describes a BPS excitation \refnew{result} vanishes.
The two possible BPS excitations are traveling waves (momentum
along the brane) and instantons (for $p \ge 4$) describing
a $p-4 $ brane inside a $p$ brane \cite{dgl}.
For a traveling wave we only have 
 $F_{- i} \not = 0$ where $ x^- = t - x^9$, and both terms in
\refnew{result} individually cancel.
For the case of the instanton let us denote by $I,J,K,L$ the
dimensions along which the gauge field is nontrivial (the
four dimensions of the instanton).
Using the self duality condition we find
\be\la{simpleinst}
F_{IJ} F^{JK  } = { 1\over 4 } \delta_{IK} F_{L J} F^{ J L}~,
 \ee
so that the two terms in \refnew{result} cancel.
We also obtain a cancellation if we have instantons and traveling
waves at the same time.

\section{Supergravity calculation and its connection to gauge theories}

We expect that if we have a very large number of D-branes 
the system will be well described by a supergravity solution.
This supergravity solution might be singular at $r=0$, but with
a bit of abuse in notation we will call it a black hole, with
the understanding that we will be careful about getting very 
close to $r=0$. 
 It was shown in \cite{jmasgrey} that the limit
\be\la{bhl} g \to 0~,~~~~~~N\to \infty~,~~~~~~~~gN ={\rm fixed} \gg 1
\ee
defines a black hole limit for a system of D-branes that is
well described by the corresponding supergravity solution. 
In this limit the string coupling is zero in the bulk so we just have
free strings in the presence of a  nontrivial background metric. 
A similar limit can be described for NS fivebranes which is 
\be\la{bhlns} g \to 0~,~~~~~~N ={\rm fixed} \gg 1
\ee
In these limits the string theory in the bulk becomes trivial.

One can define a more general black hole limit in which the 
bulk string coupling is finite. In that case one needs an 
additional low energy condition:
we take the energies to be much smaller than any other energy
scale of the theory and also we take the gravitational radius
of the black hole to be much bigger than the inverse of the
smallest energy scale of the theory. This low energy condition
ensures that the bulk theory becomes free.
In the case of ten dimensional type II string theories that
we were considering above this means that the energies should
be much smaller than $\sqrt{\alpha'}$ and the  gravitational radius should
be much bigger than $\alpha'$. We could alternatively keep the
energies
fixed and take the limit
\be\la{bhgen}
\alpha' \to 0~,~~~~~~~~~~ g \alpha'^{7-p} N = {\rm fixed } =
r_g^{7-p}
\ee

An interesting case to consider is the case of a $D3$ brane. In that
case we obtain some black hole for any value of $g$. This supergravity
solution is nonsingular, so we could put a probe arbitrarily close to 
$r=0$.

We now write the action of 
a test D-brane in the
background black hole that carries D-brane charge \cite{leigh}
\be\la{classic}
S  =  - { 1 \over g (2\pi)^p \alpha'^{p+1 \over 2}  } \int d^{p+1} x \ 
e^{-\phi}\sqrt{ det G }  + A_{01...p }~,
\ee
where $G$ is the induced metric on the brane and $A_{p+1}$ is  
 the 
$p+1$ form potential that couples to the D-brane charge. 
The background is the corresponding supergravity solution
\cite{horstrom}. 
For the extremal D-$p$-brane  solution we find 
\be\la{class} 
S = - { 1 \over g^2_{YM}\alpha'^2 } \int d^{p+1} x  
  f^{-1} \left( \sqrt{ 1 - f v^2} -1 \right)
\ee

The function $f$ is 
\be\la{deff}
f = 1 + {k\over \alpha'^2}~,~~~~~~~~~~k = { c_{7-p} g^2_{YM} N \over
(2\pi)^p  X^{7-p} } 
\ee
(remember that $X= r/\alpha'$). 

If we take the gauge theory limit that we took above, keeping $X$
fixed,
we see that the action becomes 
\be\la{claslim}
S = - { 1\over g^2_{YM} } \int d^{p+1}x  k^{-1} (\sqrt{ 1 -k \dot X^2 } -1)
\ee
where $ \dot X = v/\alpha'$ is the derivative of the field measuring transverse
displacements (note that it includes a factor of $1/\alpha'$ and has
dimensions of $1/{\rm length}^2$).
There is an interesting connection between \refnew{claslim} and the discrete
light cone quantization of \cite{susskind}.
If we compactify the space along the D-brane and we do some duality 
transformations we can transform the D-brane charge $N$ into
momentum. Then \refnew{claslim} describes the action for a 
momentum excitation moving close to the extremal  black hole carrying
momentum charge. Furthermore, the
action \refnew{claslim} is precisely the action one would obtain in 
discrete light cone quantization \cite{beckerspol}. 
%We see clearly that, in the 
%black hole case, taking the gauge theory limit leads to the
%same results as considering discrete light cone quantization. 
Notice that if we expand the action \refnew{claslim} we get a series in
powers
of $k$ which is reminiscent of \refnew{powerc}.    

Now the effective action \refnew{powerc} will be a good approximation as long
as we can neglect the creation of massive strings going between the probe
and the black hole. The mass of such massive strings is $X$ and 
the kinetic energy available is  roughly  $\omega \sim {\dot X \over X} $
therefore we can neglect massive string creation as long as
\be\la{creasup}
   \  ~~~~~~~~ { \dot X \over X^2 } \ll 1
\ee
So we see that if we keep $v^2 g^2_{YM} N/X^{7-p} $ fixed while
obeying \refnew{creasup} then from the point of view of 
supergravity we have the action \refnew{claslim} and from the
point of view the gauge theory we get \refnew{powerc}.
This limit demands that $g^2_{eff} \sim g^2_{YM} N/X^{3-p} \gg 1$,
 which is saying that
the effective large $N$ coupling at the energy scale given by $X$ is
large, so that \refnew{creasup} is obeyed and $g^2_{eff} \dot X^2/X^4 $ is 
fixed and finite. 
We should also be careful not be close 
 to creating  a near extremal black hole.
This would happen when the energy of the probe is such that
the probe lies within the Schwarzschild radius of the total system
consisting of the original extremal branes, plus the probe, plus
the energy of the probe.
This condition implies that
\be
G_N^{10} \varepsilon \ll r^{7-p}
\ee
where $\varepsilon$ is the energy density on the probe.
This implies that the following relation should hold 
\be
{ g^4_{YM} \varepsilon \over X^{7-p} } \ll 1
\ee
In our case this translates into $ g^2_{YM} \dot X^2/X^{7-p} \ll 1 $
which together with the other conditions that we found above imply
that $N \gg 1 $. Viewing the gauge theory diagrams as string diagrams
this condition implies that we consider diagrams having only one 
boundary on the probe but multiple boundaries on the background branes. 
 It is clear that
terms in \refnew{powerc} with $I> 2L +2 $ are not going to contribute while
nonzero terms with $ I < 2L +2 $ would spoil any possible agreement,
therefore
they should vanish.
The terms with $I = 2L +2 $  should have certain
specific coefficients so that the two answers match. 
In summary, the correspondence between black holes and supergravity
implies that the leading $N$ diagrams of $p+1$ dimensional gauge theories
should vanish for $I < 2 L +2 $ and should have some specific coefficients
for $I = 2L +2 $. For the case of zero branes, $p=0$, and two loops this
results were checked by \cite{beckers,beckerspol}.
 They considered the case $F_{0i} = v_i$
and
showed that the first nonvanishing term is proportional to $v^6$ 
($ I = 6 = 2 \times 2 + 2 $), which has precisely the right coefficient
to agree with supergravity. 

We can consider cases where the black hole has some excitations, which
could be BPS or not. These excitations are described by exciting the 
gauge field. We take the excitation energy to be finite in the 
gauge theory limit. This implies that the corresponding black hole
has one very large charge. 

Let us consider first BPS excitations. For simplicity, let's consider 
 the five dimensional case, $p=5$. We could have instanton-strings
and momentum running along the strings. 
The corresponding supergravity background is that of an extremal
black hole with three charges. 
The action of a probe fivebrane becomes 
\be\la{threech}
 S = { 1\over g^2_{YM} \alpha'^2 } \int f^{-1}_5
 \left( \sqrt{ 1 - f_5 f_1 f_p v^2 }
-1
\right)
\ee
where $R$ and $V_4$ are the radius of the dimension along the
string instanton and the transverse four-volume respectively and 
 $f$ is as in \refnew{deff} and 
\be\la{defone}
f_1 = 1 + { g^2_{YM} Q_1 \over V_4  X^2 } ~~~~~~~~~~~~~
f_p = 1 + { g^4_{YM} N \over R^2 V_4 X^2 }
\ee
We see that in the limit $\alpha' \to 0$ the functions \refnew{defone} remain 
finite. 
In the gauge theory limit we should replace $f$ in \refnew{threech} by
$k/\alpha'^2$ from \refnew{deff}. 
This again is the corresponding action that we would have found if we
were doing discrete light cone quantization after transforming
the fivebrane charge into momentum.
Now we could ask if \refnew{powerc} has the right structure to 
give \refnew{threech}. We will analyze this question qualitatively. 
We see that we could write $ {Q_1 \over V_4} \sim F_{inst}^2 $ 
and that $ {g^2_{YM} N \over R^2 V_4 } \sim F_{p}^2$ where
$F_{inst}$ is the typical field strength of the instanton 
configuration and $F_p$ is the typical field strength associated to
the excitations carrying momentum. 
We indeed see that the whole classical answer \refnew{threech} 
comes from the terms with $I = 2 L +2$ from \refnew{powerc}.

In the case of $p<5 $ we get results similar to \refnew{threech} but
we can only add one extra charge (breaking another supersymmetry).

We can also do a similar analysis for near extremal 
D-$p$-branes. 
In that case the black hole solution involves two functions
\be\la{nextr}\ba{rl}
f =& 1+ {k \over \alpha'^2} -{1\over 2} { r_0^{7-p} \over r^{7-p} } + 
o(\alpha'^2 /k) ~,\\
  g =& 1 - { r_0^{7-p} \over r^{7-p}}~, \\
X_0^{7-p} \equiv& { r^{7-p}_0 \over {\alpha'}^{7-p}} = { g^4_{YM} 2 (7-p)
\over
(2\pi)^p (9-p) } \varepsilon 
\ea
\ee
where $\varepsilon$ is the energy density on the brane. 
And the effective action is 
\be\la{nexclass}\ba{l}
S = -{1\over g^2_{YM}\alpha'^2} \int d^{p+1}x  \times 
\\ ~~~~~~~~~f^{-1} \left[
\sqrt{ h - f ( {v^2 \over h}+ r^2 \dot \Omega^2 ) } +
{k }  \right] \ea
\ee
In the gauge theory limit this action becomes
\be\la{nexdlcq}\ba{l}
S = - {1\over g^2_{YM}} \int d^{p+1}x \times\\~~~~~~~~
 k^{-1} \left[
\sqrt{ h - k ( {\dot X^2 \over h } + X^2 \dot \Omega^2 ) } +
 k   \right] ~.  \ea
\ee
Expanding \refnew{nexdlcq} we obtain an expression of the
form \refnew{powerc} after identifying $g^2_{YM} \varepsilon  \sim F^2 $. 
Again the terms with $I = 2 L +2 $ are expected to contribute.
Actually when we compare the supergravity and the gauge theory 
we might need to make a change of variables. This change of variables
is not arbitrary, it will be determined by the relative coefficients
of the terms proportional to $\dot X^2$ and $X^2 \dot \Omega^2$.

\subsection{ Connection with string theory diagrams}

\footnote{I am indebted to  M. Douglas and S. Shenker for discussions on
this section.}
The series in \refnew{powerc} can be viewed as
the $X \to 0 $ limit of some string diagrams. These are diagrams
having $L + 1 $ holes, one of the holes is attached to 
the probe D-brane and the $L$ others are attached to the black hole. 
It is easy analyze the  $X \to \infty$ limit of the diagrams, when
 the branes are very widely separated, since it 
 reduces to a closed string tree level diagram.
This tree level diagram, of course, gives  the supergravity
result,
since the classical low energy action sums up all the tree level diagrams.
Therefore for $ X \to \infty$ the string diagram agrees with the
supergravity calculation. The non-trivial fact is that the 
string diagram should also agree for $X \to 0$. This seems to imply that
the string diagram is the same for short and long distances (apart from
the trivial scaling with $X$ 
already present in \refnew{powerc}). In principle,
the diagram could involve a nontrivial function  $ F(r^2/\alpha')$. 
It seems that there could be a non-renormalization argument
(in $\alpha'$)
explaining this, but it is unknown to me. 
An important fact is that the results are expected to agree with
supergravity
only after performing an average over all possible microscopic
configurations
with given charges and masses. We will see an example of
this for a one loop near extremal calculation. 

\subsection{Nonperturbative corrections}

The in the above subsection we indicated that we expect to find agreement
order by order in perturbation theory in $g$ between the gauge theory
diagrams
and the long distance supergravity calculation. This implies some
perturbative
non renormalization theorems that forbid terms with $I <L+2$ (at least for
large $N$). In \cite{jppp,dineseiberg} this question was addressed 
for the $I=4$ ($F^4$) terms. The conclusion was that for $ p=3$ they were
not renormalized while for $p=2$ there were nonperturbative corrections.
These corrections are in fact necessary for obtaining type IIB string
theory
from matrix theory \cite{twobmatrix}. These terms arise naturally in the
study of the D2 brane black hole solution. This solution has the feature
that
the dilaton blows up as we approach the core. This implies that we should
use the 11 dimensional supergravity description. For large $N$ the
curvatures will be small in the classical solution. From the 11 dimensional
point of view we have M-twobranes perpendicular to $R_{11}$. Since there is
a no force condition we should decide how we distribute the M-twobranes on
this circle. If we just ``lift -up'' the ten dimensional D2-brane solution
we obtain a solution in 11 dimensions which would correspond to 
M2-branes uniformly distributed along $R_{11}$. If we add a very small
amount of energy this solution would collapse into one in which the
branes are all sitting at a point in $R_{11}$\footnote{
If the energy we add is large enough then the solution in which they 
are distributed uniformly becomes the stable one \cite{mynotes}.}.
The metric of the M-twobrane solution is
\be \la{mtwo}
ds_{11}^2 = H^{-2/3} dx_{||}^2 + H^{1/3} dx^2_{\perp} 
\ee
where $H$ is a harmonic function and $x_{||},~x_{\perp}$ indicates the
coordinates
parallel and perpendicular to the brane respectively.
In the case that the branes are all sitting a point in $R_{11}$  the
harmonic function is 
\be \la{harm}
H = 1 + \sum_n { c_6 l_p^6 N \over | x - n \vec{R}_{11} |^6 }
\ee 
where $l_p = g^{1/3} \sqrt{\alpha'}$. For large $x$ \refnew{harm} becomes 
\be \la{dtwo}
H \sim  1 + { c_5 g^2_{YM} N  \over ( 2 \pi)^2 \alpha'^2 X^5 }
\ee
and the change occurs when $ r \sim R_{11} $ or in other words when
$X \sim  {R_{11} \over \alpha'} \sim  g^2_{YM} $.
 It is precisely for energy scales lower
that
the coupling constant of the gauge theory that we expect to start flowing
into
the strong coupling infrared fixed point. Notice also that 
the physical value of $R_{11}(r) = H^{1/6} R_{11} $ is of the order
of $N^{1/3} l_p$ at $r = R_{11}$, in other words, the physical
value of $R_{11}$ at the point where the transition happens is much
larger that the Planck length when $N$ is large so that we can trust the
classical solution. 
It can be seen that for $r \ll R_{11}$  we can rescale the fields 
($ Y = X/g_{YM}$) so that
$g_{YM}$ disappears from the action and the transverse displacement 
fields, $Y$,  have  dimension 1/2, if we are very close to the
core of the 11 dimensional M-twobrane solution, $r \ll  R_{11}$, 
the fact that $R_{11}$ is compact becomes irrelevant
  and the action becomes SO(8) invariant.

This is in agreement with the expectations from matrix theory since
in the limit that the coupling is very large we expect to be describing 
a type IIB string theory in ten dimensions \cite{twobmatrix}.
The non-perturbative corrections are expected to precisely change the
form of the Harmonic function from \refnew{dtwo} to \refnew{harm}. This is
easy to see by using the Poisson resumation formula in \refnew{harm}. The
necessary exponential corrections go as powers of 
$e^{- X/g^2_{YM} } \sim e^{-r/R_{11} } $. In fact the $v^4$ terms were
calculated in \cite{jppp} for the one instanton contribution and
summed for all instantons  in \cite{multi}. 

In summary, when the  transverse displacement field is such that its
value is bigger than the energy scale associated to the Yang-Mills coupling
of the 2+1 gauge theory we have the effective action expected for
D-twobranes in string theory. When the transverse displacement field 
has an expectation value comparable to the gauge theory coupling
instanton corrections become important and change the behavior such that
for expectation values much smaller than the energy scale of the theory
we get the conformal fixed point behavior, 
which is the SO(8) invariant 
behavior of an M-twobrane moving in the presence of $N$
M-twobranes.

The same kind of transition is expected to happen for the IIA theory NS 
fivebrane. In that case the transition will happen when the expectation
value of the field measuring transverse displacements becomes
of order $X \sim 1/ \sqrt{\alpha'} $ where $\alpha'$ is the energy scale
that characterizes the decoupled IIA fivebrane theory. 
For lower values of $X$ we start seeing the 11 dimensional character of the
solution and we expect to find an enhanced SO(5) symmetry, characteristic
of
the (0,2) fixed point theory describing M-fivebranes in 11 dimensions. 
This is of course describing the limit in which one of the radii of $T^5$ of
the original M-theory is going to infinity \cite{seibergfive}.

\section{ Some one loop results}

We summarize some one loop results that show how the different cases
mentioned above work.

\subsection{Calculation of the $v^2$ forces} 

Consider a near extremal black hole. A probe moving in its presence 
will feel a static force (proportional to $v^0$) and also a force 
proportional to $v^2$. We calculate here the $v^2$ force, leaving the other
for the next subsection. Instead of taking the probe to be moving we could,
of course, take the black hole to be moving. 
This term can be calculated by inserting  in \refnew{result} the field strength
$F = v_{0i} + \tilde F $ where $\tilde F$ describes the excitations on the
brane that raise the brane  energy above extremality.

\be\la{twovel}\ba{rl}
S_2 =&  { c_{7-p}  \over(2\pi)^p   X^{7-p} } 
\int d^{1+p } x \ \left [ {\dot X^2 \over 2 } \ g^2_{YM} \ T_{00} 
- \right.\\  & \left. { 1 \over 2  }  Tr [ F_{1 i \mu } F^{ ~~\mu }_{1 j} ]
\dot X^i \dot X^j 
+ {\rm fermions} \right ] ~,
\ea \ee
where 
\be\la{tzz} \ba{rl}
T_{00} =&  { 1 \over g^2_{YM} } \left\{ Tr [F_{1\mu 0} 
F_{1~ 0}^{~ \mu} ] \right. + \\ 
& \left. { 1\over 4 }  Tr [F_{1\mu \nu } F_{1}^{~ \mu \nu } ]   
+ {\rm fermions} \right\}
\ea \ee
is the stress tensor associated to the unbroken $U(N)$ Yang-Mills 
theory.
Supersymmetry implies that the fermions will appear in \refnew{tzz}
contributing
to the stress tensor. 

We are interested  in the case where the velocity is
along the directions transverse to the branes. We are just saying   that
the probe Wilson lines do not have a constant time dependence
($ F_{2 I 0} = 0$), in other words: there are no
winding fundamental strings dissolved on the probe. We are also interested
in evaluating \refnew{twovel} in an average sense.
Under these
 conditions the term proportional to $\dot X^i\dot X^j$ in \refnew{twovel}
does not contribute since it will be proportional to $ D_\mu X^i D^\mu
X^j $, we can integrate by parts the covariant derivative and 
then the term vanishes using the equations of motion. 
This implies that \refnew{twovel} becomes 
\be\la{twovelfin}
S_2 = { c_{7-p} g^2_{YM} \over (2\pi)^p  X^{7-p} } 
  \int d^{p+1}x  {\dot{ \vec{ X}}^2 \over 2 } \  \varepsilon~,
\ee
where $\varepsilon$ is the  energy density on the D-brane. 
Now we compare this with the supergravity prediction. 
Expanding \refnew{nexdlcq} in powers of the velocity we find a static 
potential, a $v^2$ term, etc.
We start calculating the $v^2$ term 
\be\la{vsquare} { 1 \over g^2_{YM} } \int d^{p+1}x {1\over 2} 
{ \alpha' \over \sqrt{h}} \left( { \dot r^2 \over h } + 
r^2 \dot \Omega^2
\right)
\ee
We notice that 
the
coefficient is different for $\dot X^2 $ and $X^2 \dot \Omega^2$ while
in our result \refnew{twovelfin} the coefficient was the same.
The resolution to this apparent contradiction is that the coordinate
$r$ of the spacetime solution is not necessarily the same as the 
coordinate $r$ of the D-brane calculation.
Let us define a new radial coordinate $\rho$ by the equation
\be \la{rhoeq}
{ d \rho \over \rho } = { dr \over r \sqrt{h} } ~ .
\ee 
Using \refnew{rhoeq} the  term \refnew{vsquare} becomes 
proportional to $ v_\rho^2 = ( \dot \rho^2 + \rho^2\dot \Omega^2 )$,
so that 
$\rho$ is interpreted as the distance that appears in the
D-brane calculation, with $X = \rho/\alpha'$.
It is interesting that according to the new variable  the position 
of the branes is at $\rho =0$ or $r= r_0$ which is the horizon.
This
 comment should not
be taken too seriously because we have only calculated the one loop
term
and, in principle, to get close to the horizon it would be necessary to
calculate higher loops.
Expanding \refnew{class} to leading order in $r_0^{7-p}/r^{7-p}$, taking 
into account the change of coordinates \refnew{rhoeq} we get the 
velocity dependent term
\be\la{velocact}
S = { 1 \over g^2_{YM} } \int d^{p+1}x  
{ \dot{ \vec{X}}^2 \over 2 } ( 1 + { (9-p) \over
2 (7-p)} { X_0^{7-p} \over X^{7-p} } )~.
\ee 
Expressing $r_0^{7-p}$ in terms of the energy \refnew{nextr} we find
again
\refnew{twovelfin} , in precise agreement with the D-brane probe calculation.
Notice that it was crucial to perform a change of
variables to find 
 agreement.
This quantity agrees for any $p$, this is due 
to the simplicity of the operator that couples to $v^2$: it is just
the
physical energy, so it is not renormalized by strong coupling effects
(large $g N$ effects).

The result \refnew{twovelfin} 
 includes, 
as a particular case, the one loop calculation of
\cite{dps}. In that case,
$p=5$ and  we have  an extremal state containing $Q_1$ instanton strings
and momentum $N$. The total  energy on the fivebrane worldvolume is 
\cite{hms} 
\be\la{energy}
E = {N\over R } + { R Q_1 \over g \alpha' }~,
\ee
so that the probe metric becomes 
\be\la{twofivedim}\ba{rl}
S=& {R V  \over g} \int dt {v^2 \over 2 } \left( 1 + 
{g^2 \alpha'^4 N \over R^2 V r^2 }\right.
+  \\   + & \left. { g \alpha'^3  Q_1 \over V r^2 } + 
{ g^3 \alpha'^7 N Q_1 \over R^2 V^2 r^4 }   \right)~,
\ea \ee
where the $o(g^3/r^4)$ term should come from a two loop calculation
\cite{dps}. Indeed we see form the arguments given below \refnew{threech}
that \refnew{powerc} has the right structure to produce this term. 
It is a question of performing the detailed calculation of the coefficient
to see that they match.

\subsection{Calculation of the static 
force}

The calculation of the one loop contribution to the static force reduces to
evaluating \refnew{result}      
in some generic thermal ensemble. 
This is  difficult in principle  because we have not calculated the 
fermionic terms and they will contribute. We will do the calculation
in a case where it is easy to see what the effect of the fermions is.
Of course, as explained in \cite{ascv}  the supergravity solution
is expected to agree only in the limit of large $g N $. Which means
that the effective large $N$ coupling of the gauge theory is strong.

We will calculate \refnew{result} in a configuration carrying the charges 
of the 
the five dimensional near extremal black hole of \cite{cama,ghas} in the
dilute gas regime. 
Then  $p=5$,  we call $N = Q_5$ and
we also put $Q_1$ instanton strings along one of the directions
of the fivebrane, let us call it the direction $\hat 9$. Even though
they are called ``instantons'' these objects are physically string
solitons
of the 1+5 dimensional gauge theory.
The instanton configuration is characterized by some moduli $\xi^r$,
$r = 1,.., 4Q_1Q_5$.
These moduli can oscillate when we move along the direction $\hat 9$,
$\xi(t,x_9)$. We are interested in the case where the energy of the
oscillations is small, so that we can describe the excitations
of the system as oscillations in moduli space. 
The condition  is that the total energy due to the
oscillations
should be much smaller than the energy of the instantons $ E \ll R_9
Q_1/g \alpha' $.
(In the notation of \cite{hms,jmasgrey} this means $r_n \ll r_1 $.)
This picture of the D-1-brane charge being carried by instantons
in the gauge theory is correct when the energy of the instantons
is much smaller that the total  mass of the fivebranes 
$ M_1 = R_9 Q_1/g\alpha' \ll M_5 =  R_5R_6R_7R_8R_9 Q_5/g\alpha'^3 $ 
(this  means $r_1 \ll r_5$).
So that we are in the dilute gas regime of \cite{jmasgrey} \footnote{
The definition of the dilute gas regime $r_n \ll r_1 , r_5 $ does
not require any specific relation between $r_1$ and $r_5 $.
}.
Calling $x^\pm = x^9 \pm t$,  the nonzero components of the gauge
field are $F_{\pm I} = \partial_\pm \xi^r \partial_r A_I $ with 
$I = 5,6,7,8$ and $F_{IJ}$,  the field of the instanton.
The action for the small  fluctuations of the instanton configuration
becomes
\be\la{actionfl}\ba{rl}
S_0 =& 
{ 1 \over g (2\pi)^5 \alpha' }
 \int d^{1+5} x  \ { 1\over 2} Tr[ F_{\alpha I } F^{\alpha I} ]
+~~
{\rm fermions } \\ = & { 1\over 2 } \int dt dx^9 
G_{rs}(\xi) \partial_\alpha
\xi^r
 \partial^\alpha \xi^s +~~
{\rm fermions }~,
\ea \ee
where $\alpha = \pm $.
We  have an nonlinear sigma model 
action for the instanton fluctuations \cite{ascv}. 
The theory \refnew{actionfl} has (4,4) supersymmetry and the metric
$G_{rs}$ is hyperk\"ahler \cite{ascv}. 

Using \refnew{simpleinst} 
we can see that the effective action \refnew{result} reduces to 
\be\la{potential}\ba{rl}
S_1 = &{ 2 \over  ( 2\pi )^5  X^{2} }  
\int d^{1+5 } x \  \ Tr [ F_{+I} F_{+ I} F_{-J}
F_{-J }] + \\ ~~~~~~~~ &+
{\rm fermions }~.
\ea \ee
When we integrate over $t,x^9$ we will effectively average 
separately the term with $++$ and the ones proportional to $--$.
We assume that the oscillations average the fields in such a way
that we can replace $ F_{+I} F_{+I} $ by its average value, both 
in spacetime indices and group indices. So we get
\be\la{potentialnew} \ba{rl}
S_1 = &{ 2  \over  ( 2\pi )^5  X^{2} } 
\int dt \int d^{5 } x \ Tr [ F_{+I} F_{+ I}] \times \\
& ~~\times { 1\over ( 2\pi)^5 R V Q_5 } \int d^5 x Tr[ F_{-J} 
F_{-J }] 
\\
=& { 2 g^2 \alpha'^2 \over R V Q_5 } {1\over X^2}
 \int dt \int dx^9 T_{++} \int dx^9  T_{--} \\
=& { 2 g^2 \alpha'^2  E_L E_R \over R V Q_5 } {1\over X^2}\int dt 
\ea \ee
where $R$ is the radius of the $9^{th}$ direction and $V =R_5R_6
R_7R_8$
is  the product of the radii in the other four  directions. 
$E_{L,R} $ are the left and right moving energies of the 
effective two dimensional theory  \refnew{actionfl}.
Notice that we have calculated only the bosonic terms. Supersymmetry
then implies that the fermions appear in \refnew{potentialnew} just
as another energy contribution.
The form of the operator in  \refnew{potentialnew} is the identical to the
one that appeared in the calculation of the
fixed scalars grey-body factors \cite{cgkt}.

The supergravity calculation \refnew{nexdlcq} predicts the static potential
\be \la{static}
V = { 1 \over g^2_{YM} } k^{-1} (\sqrt{h} -1 )
\ee
Expanding this to first order in $1/r^{7-p} $ we find the one loop
contribution.
Expressing it in terms of the parameters of the five dimensional black hole
that we were considering above we see that it 
exactly matches \refnew{potentialnew}.

In this case we see that if we do not take a typical configuration and
average
as we did above we would not get agreement with supergravity. For example,
we
might take a left moving wave localized at some point in the internal $T^4$
and
a right moving one 
 localized at a different point of the internal $T^4$. For these
excitations \refnew{result} would vanish. Of course these configurations do not
quite solve the equations of motion, so one can hope that the equations of
motion of the excitations along the brane are
ergodic, so that the time average is equivalent to a thermodynamic
average.

\section{ Concluding remarks}

We saw how supergravity solutions demand 
a certain behavior for large $N$ diagrams in gauge theories.
We found that the subleading corrections would be small when we can
neglect the open strings stretching between the probe and the black hole. 
This correspondence between gauge theory results and supergravity 
results  arises just from the physics of black holes, but can, of course, 
be of use in matrix theory. 
We have  checked here  this  correspondence to one loop in a wide
variety of situations, including near extremal black holes.
In \cite{beckers,beckerspol}
 a two loop calculation was done for a special case.
 We  conjecture that the results will
agree up to an arbitrary number of loops. Furthermore we  conjecture
that the large $N$ diagrams that are needed are finite for $ p =4,5,6$.
If they were  not finite,  they should be calculated in the 
nontrivial (0,2) theory (for $p=4$) or the the NS fivebrane theory
for $p=5$. But the definition of these theories themselves as the theories
of coincident branes implies that any formulation of the theory, such
as the one proposed in \cite{ofereva},  should be 
such as to provide agreement for these terms. 
All these  are the terms needed to show that matrix
theory incorporates correctly all nonlinear classical 
 supergravity effects. 
It is interesting to notice that the nonlinear form of the action 
\refnew{claslim}  
is determined by local Lorentz invariance. It would be interesting
to know what this principle translates into for the gauge theories. 
The connection with the string diagrams suggests that these
string diagrams should be independent of $\alpha'$ (apart from trivial 
factors), so in some sense they are ``topological''.
This has been verified explicitly for the one loop diagram  with
four insertions in \cite{dkps}. This also implies many perturbative 
nonrenormalization
theorems protecting the diagrams less than $2L +2$ insertions. In principle
one might only need a large $N$ non-renormalization theorem.
We expect nonperturbative corrections which have a clear physical
interpretation
for the 
$p =2$ case \cite{dineseiberg}.
These corrections are  saying that a black
hole made with D2 branes, when approached closely enough will start having
an 11 dimensional character, since the D2 branes become M-2-branes
localized
along $R_{11}$ and the local size of $R_{11}$ is growing as we approach 
the branes since the dilaton diverges for a D-2-brane. 
This of course has a well known matrix theory interpretation as the
IIB limit of M theory on $T^2$. 

This correspondence between large $N$ gauge theories and supergravity 
will most probably provide new insights both for gravity and gauge large
$N$ gauge theories that are only beginning to be explored.

{\bf Acknowledgments}

I very grateful to V. Balasubramanian, T. Banks, M. Douglas,
F. Larsen,  N. Seiberg,
S. Shenker, A. Strominger  and C. Vafa 
for interesting  discussions.
This work was  supported in part by 
DOE grant
DE-FG02-96ER40559. It is a pleasure to thank the
organizers of Strings97 for a wonderful conference.

\end{document}